\documentclass[12pt]{article}

\usepackage[english]{babel}
\usepackage[cp1251]{inputenc}
\usepackage[T1]{fontenc}

\pdfoutput=1
\usepackage{makeidx}
\usepackage{amssymb}
\usepackage{amsfonts}
\usepackage{amsmath}
\usepackage{mathrsfs}
\usepackage{graphicx}
\usepackage{setspace}
\usepackage{authblk}
\usepackage{units}
\usepackage{appendix}
\usepackage{xparse}
\usepackage{cite}
\usepackage{hyperref}
\usepackage[left=2.1cm,top=2.5cm,right=2.1cm]{geometry}

\begin{document}
	
\title{Non-commutative integration method and generalized coherent states}
\author{A. I. Breev$^{1}$\thanks{
		breev@izmiran.ru}, D. M. Gitman$^{2,3}$\thanks{
		dmitrygitman@hotmail.com}
	, \\
	$^{1}$\small{Pushkov Institute of Terrestrial Magnetism, 
		Ionosphere and Radiowave Propagation (IZMIRAN), \\
		4 Kaluzhskoe shosse, 108840 Troitsk, Moscow, Russia;}\\
	$^{2}$ P.N. Lebedev Physical Institute, \\
	53 Leninskiy ave., 119991 Moscow, Russia.\\
	$^{3}$ Institute of Physics, University of S\~{a}o Paulo, \\
	Rua do Mat\~{a}o, 1371, CEP 05508-090, S\~{a}o Paulo, SP, Brazil.}

\maketitle

\begin{abstract}
	The relationship between states obtained by the non-commutative integration method of the Schr\"{o}dinger equation on Lie groups and generalized coherent states is investigated. It is shown that such solutions belong to the class of generalized coherent states when the corresponding $\lambda$-representation is real.
	%
    %
\end{abstract}

\section{Introduction}

Coherent states first arose in quantum mechanics as solutions of the Schr\"{o}dinger equation that minimize the Heisenberg uncertainty relation (see Refs. \cite{1_Charp,7_Manko} and references therein). Generalizations of these states find wide application in various areas of physics \cite{2_coh_opt,3_coh_inf,4_coh_inf_2}.

Among the generalizations of coherent states, one can distinguish nonlinear coherent states \cite{5_non_coh}, Gazeau-Klauder coherent states \cite{6_GK_coh}, Barut-Ghirardello coherent states \cite{7_BG_coh}, and Perelomov coherent states  \cite{8_perelomov}. When constructing Prelomov coherent states, a group-theoretical approach is used. It turns out that using the representation of the Heisenberg-Weyl group in an arbitrary Hilbert space, one can reproduce the coherent states of a harmonic oscillator. This observation allows one to construct a wide class of states using representations of various Lie groups. It is also worth noting that the concept of coherent states can be generalized to quantum Lie groups \cite{8_perelomov}.

In this paper, we study the stationary Schr\"{o}dinger equation on Lie groups from the point of view of constructing Perelomov generalized coherent states and the method of non-commutative integration of linear differential equations (NI), which was proposed in the works \cite{9_Sh_Shir,darbu}. This method essentially uses the symmetry of a differential equation and its algebra of symmetry operators and allows one to construct a basis of solutions that, in general, differs from solutions constructed by separation of variables and from coherent states. The NI has been effectively used to construct exact solutions to the Schr\"{o}dinger, Klein-Gordon \cite{bar}, and Dirac \cite{br20,br16,br14} equations. Note that NI is closely related to the Kirillov-Konstant orbit method \cite{10_Kirill}, which allows one to construct representations of Lie groups (see Refs. \cite{Aldaya1,Aldaya2,Aldaya3,Aldaya4,Aldaya5}). It turns out that the basis of solutions that can be obtained by NI is similar to Perelomov generalized coherent states.

The paper is organized as follows. Section 2 introduces the basic definitions of Lie group and algebra theory, as well as the definition of Perelomov generalized coherent states on Lie groups. Section 3 is devoted to the construction of a special irreducible representation of Lie groups, which is a necessary construction for NI. Section 4 considers the non-commutative reduction of the Schr\"{o}dinger equation on a Lie group from the perspective of the method for constructing coherent states. Section 5 discusses the relationship between spin coherent states and states obtained using the non-commutative integration method on the rotation group $\mathbb{SO}(3)$. Section 6 contains some concluding remarks.

\section{Generalized coherent states on Lie groups}

First, we introduce some facts from Lie group theory necessary for
the further exposition.

Let $G$ be an $n$-dimensional Lie group. The tangent space $T_{e}G$ at the identity $e\in G$ forms the Lie algebra $\mathfrak{G}$ of the Lie group $G$. Let $\left\{ e_{A}\right\} $ be some fixed basis in $\mathfrak{G}$ ($A=1,\dots,n$),
\begin{equation}
	[e_{A},e_{B}]=C_{AB}^{C}e_{C},\quad A,B,C=1,\ldots,n,\label{Lie_algebra}
\end{equation}
where $C_{AB}^{C}$ are the structure constants of the Lie algebra $\mathfrak{G}$.
Any element $X\mathfrak{\in G}$ can be uniquely factored over this basis, $X=X^{A}e_{A}$. The Lie group acts on itself by right $R_{g}$ and left $L_{g}$ shifts ($R_{g'}g=gg'$, $L_{g'}g=g'{}^{-1}g$), the differentials $(R_{g})_{*}$ and $(L_{g})_{*}$ of which generate left- and right-invariant vector fields on the group $G$, respectively:
\begin{align*}
	& \xi_{X}(g)=(L_{g})_{*}X,\quad\eta_{X}(g)=-(R_{g})_{*}X,\quad\left[\xi_{X}(g),\eta_{Y}(g)\right]=0,\\
	& \left[\xi_{X}(g),\xi_{Y}(g)\right]=\xi_{[X,Y]}(g),\quad\left[\eta_{X}(g),\eta_{Y}(g)\right]=\eta_{[X,Y]}(g),\quad X,Y\in\mathfrak{G}.
\end{align*}
Left-invariant and right-invariant vector fields are the generators of the right $T^{R}(g)$ and left $T^{L}(g)$ regular representation of the group:
\begin{align}
	& \xi_{X}(g)f(g)=\left.\frac{d}{dt}\right|_{t=0}T^{R}\left(\exp(tX)\right)f(g),\quad T^{R}(g')f(g)=f(gg'),\label{eq:Left_Right}\\
	& \eta_{X}(g)f(g)=\left.\frac{d}{dt}\right|_{t=0}T^{L}\left(\exp(tX)\right)f(g),\quad T^{L}(g')f(g)=f(g'{}^{-1}g),
\end{align}
where $f=f(g)$ is a function on the group, $g,g'\in G$, $\exp:\mathfrak{G}\rightarrow G$ is the exponential map.

Generalizing the definition of a group action to manifolds of lower dimension, we arrive at the definition of a homogeneous space. Let $G$ be an action of a group on an $m$-dimensional manifold $M$ using a smooth function $M\times G\rightarrow M$, $x'=x\circ g$, $x\in M$. Suppose that $G$ acts transitively on $M$. In this case, $M$ is a homogeneous space of $G$ and $m\leq n$.

Let us describe how a homogeneous space is structured locally. Let $H$ be the isotropy Lie subgroup of a point $x_{0}\in M$ and $\dim{H}=n-m$, then $M\simeq G/H$, where $G/H$ is the space of right cosets. This interpretation allows us to consider the Lie group $G$ as a principal bundle with canonical projection $\pi:G\rightarrow M(\simeq G/H)$ and structure group $H$. Let $s:M\rightarrow G$ be a section of the bundle $(G,\pi,M,H)$, then any element of $G$ can be represented as \cite{3_Nakahara}: 
\begin{equation}
	g=hs(x),\quad h\in H,x\in M.\label{Hom_g}
\end{equation}
Based on the idea of $M$ as a $G/H$ space, we can write: 
\begin{equation}
	s(x)g=h(x,g)s(x\circ g),\label{fackthomm}
\end{equation}
where $h(x,g)$ is the \textit{homogeneous space factor}.

Let us consider the application of the method of coherent states for the case when the Hamiltonian $\mathscr{H}$ of a given quantum system admits a Lie group $G$ of dynamical symmetry. 

One of the important cases for physical applications is when the Hamiltonian $\mathscr{H}$ is a quadratic form of right-invariant vector fields $\eta_{A}(g)=\eta_{e_{A}}(g)$:
\begin{equation}
	\mathscr{H}=c^{AB}\eta_{A}\eta_{B}+c^{A}\eta_{A},\quad A,B=1,\dots,\mathrm{dim}G.\label{1.3}
\end{equation}
In this case, the left-invariant vector fields $\xi_{X}(g)$ are integrals of motion of the given quantum system. If the symmetric matrix $c^{AB}$ is non-singular, then we say that the group $G$ is defined by a right-invariant metric $c^{AB}e_{A}\otimes e_{B}$.

The regular representations $T^{L/R}(g)$ of $G$ act in the Hilbert space $\mathfrak{H}$ of states of a system with Hamiltonian (\ref{1.3}). Elements of the Hilbert space $\mathfrak{H}$ will be denoted by ket vectors, $\left|\psi\right\rangle \in\mathfrak{H}$.

The Hamiltonian is invariant under the operators $T^{R}(g)$ of the right regular representation, while under the action of the operators $T^{L}(g)$ the right-invariant vector fields are transformed according to the adjoint representation $\mathrm{Ad}_{g}: \mathfrak{G\rightarrow\mathfrak{G}}$ of the group $G$:
\begin{align}
	& T^{L}(g')\eta_{X}(g)T^{L}(g'{}^{-1})=\eta_{\mathrm{Ad}_{g'}X}(g),\nonumber \\
	& \mathrm{Ad}_{g}X=\left(R_{g^{-1}}L_{g^{-1}}\right)_{*}X.\label{1.4}
\end{align}
As a result of the transformation (\ref{1.4}), the Hamiltonian (\ref{1.3}) is transformed as follows:
\begin{align*}
	& \widetilde{\mathscr{H}}=T^{L}(g')H(g)T^{L}(g'{}^{-1})=c^{AB}(g')\eta_{A}\eta_{B}+c^{A}(g')\eta_{A},\\
	& c^{AB}(g')=c^{CD}\left(\mathrm{Ad}_{g'}\right)_{C}^{A}\left(\mathrm{Ad}_{g'}\right)_{D}^{B},\quad c^{A}(g')=c^{B}\left(\mathrm{Ad}_{g'}\right)_{B}^{A},
\end{align*}
where $\left(\mathrm{Ad}_{g'}\right)_{B}^{A}=\left(\mathrm{Ad}_{g'}e_{B}\right)^{A}$ is the matrix of the adjoint representation with respect to a fixed basis in the Lie algebra $\mathfrak{G}$, $\mathrm{Ad}_{g'}X=X^{A}\left(\mathrm{Ad}_{g'}\right)_{A}^{B}e_{B}$. 

In this paper, for the convenience of further exposition, we will assume that the left-invariant Haar measure $d\mu(g)$ on the Lie group $G$ coincides with the right-invariant Haar measure $d\mu(g^{-1})$, that is, the group is unimodular ($\vert\mathrm{det}\mathrm{Ad_{g}}\vert=1)$. In the Hilbert space $\mathfrak{H}$, the scalar product is defined
\begin{equation}
	\langle\psi_{1}\vert\psi_{2}\rangle=\int_{G}\overline{\psi_{1}(g)}\psi_{2}(g)d\mu(g),\quad\psi_{k}(g)=\langle g\vert\psi_{k}\rangle.\label{1.3b}
\end{equation}
So $\mathfrak{H}=L_{2}(G,d\mu(g))$.

Following \cite{8_perelomov}, the system $\{\left|\overline{g}\right\rangle\}_{g\in G}$, where $\left|\overline{g}\right\rangle=T^{L}(g^{-1})\left|\overline{0}\right\rangle$ for a fixed vector $\left|\overline{0}\right\rangle \in\mathfrak{H}$ will be called a generalized system of coherent states for the Lie group $G$.

Suppose that $H\subset G$ is a subgroup for which the property 
\[
T^{L}(h^{-1})\left|\overline{0}\right\rangle=e^{-i\gamma(h)}\left|\overline{0}\right\rangle,\quad h\in H,
\]
where $\gamma(h)\in\mathbb{R}$ holds. Since states differing by a phase factor define a single state, the subgroup $H$ is the stationarity group of the state.

Since $T^{L}$ is a representation, the property
\[
\gamma(h_{1}h_{2})=\gamma(h_{1})+\gamma(h_{2})
\]
for $h_{1},h_{2}\in H$ holds, i.e., the function $\gamma:H\rightarrow\mathbb{R}$ is a one-dimensional representation of the subgroup $H$. Furthermore, it is clear that the system of coherent states is defined up to a subgroup $H$; namely, all information is contained in the homogeneous space $M\simeq G/H$. By (\ref{Hom_g}), for $g\in G$ we have
\begin{equation}
	T^{L}(g^{-1})\left|\overline{0}\right\rangle=T^{L}((hs(x))^{-1})\left|\overline{0}\right\rangle=e^{-i\gamma(h)}\left|\overline{s(x)}\right\rangle.\label{coh_P}
\end{equation}
This expression defines an arbitrary generalized coherent state. Finally, consider the action of $T^{L}(g'),~g'\in G$ on the coherent state $|s(x)\rangle$:
\begin{align}
	T^{L}(g')\left|\overline{s(x)}\right\rangle & =T^{L}((s(x)g')^{-1})\left|\overline{0}\right\rangle\label{Perelomov_coh}\\
	& =T^{L}((h(x,g)s(x\circ g'))^{-1})\left|\overline{0}\right\rangle=e^{-i\gamma(h(x,g'))}\left|\overline{s(x\circ g')}\right\rangle.\nonumber
\end{align}
Here we used (\ref{Hom_g},\ref{fackthomm}). The expression (\ref{Perelomov_coh}) shows that when the group representation operator acts on a coherent state, a coherent state is again obtained. 

\section{$\lambda$-representation of a Lie group\label{part2}}

The dual space $\mathfrak{G}^{*}$ of the Lie algebra $\mathfrak{G}$ is the linear space of linear functionals on $\mathfrak{G}$. Let $\left\{ e^{A}\right\} $ be some fixed basis in $\mathfrak{G^{*}}$ such that $\langle e^{A},e_{B}\rangle=\delta_{B}^{A}$, where $\langle\cdot,\cdot\rangle$ is the natural pairing of the spaces $\mathfrak{G}$ and $\mathfrak{G}^{*}$. Then the linear functional $f=f_{A}e^{A}\in\mathfrak{G}^{*}$ acts on the vector $X$ according to the rule $f(X)=\langle f,X\rangle=X^{A}f_{A}$. The Lie group $G$ acts on the Lie algebra $\mathfrak{G}$ via the adjoint representation $\mathrm{Ad}_{g}$. In turn, we can define the coadjoint representation $\mathrm{Ad}_{g}^{*}:\mathfrak{G^{*}}\rightarrow\mathfrak{G}^{*}$ by the rule
\begin{equation}
	\langle\mathrm{Ad}_{g}^{*}f,X\rangle=\langle f,\mathrm{Ad}_{g^{-1}}X\rangle,\quad f\in\mathfrak{G}^{*},\quad X\in\mathfrak{G}.\label{Adconj}
\end{equation}
Under the action of the coadjoint representation, the space $\mathfrak{G}^{*}$ is partitioned into orbits of the coadjoint representation (K-orbits). For any functions $\Phi_{1},\Phi_{2}\in C^{\infty}(\mathfrak{G}^{*})$, the Poisson-Lie bracket is defined
\begin{equation}
	\{\Phi_{1},\Phi_{2}\}=C_{AB}^{C}f_{C}\frac{\partial\Phi_{1}(f)}{\partial f_{A}}\frac{\partial\Phi_{2}(f)}{\partial f_{B}},\quad f=f_{A}e^{A}.\label{PLie}
\end{equation}
Functions $K_{\mu}(f)$ that commute with all functions on $\mathfrak{G^{*}}$ with respect to (\ref{PLie}), are called Casimir functions. The number of independent Casimir functions is called the index of the Lie algebra $\mathfrak{G}$ and is denoted by ind$\mathfrak{G}$.

Denote by $O_{\lambda}$ the K-orbit passing through the functional $\lambda\in\mathfrak{G}^{*}$. On each orbit, the Kirillov-Konstant symplectic form $\omega_{\lambda}$ is defined (see Ref. \cite{10_Kirill}). Denote by $M_{(s)}$ the subspaces invariant under the coadjoint action, which are the union of orbits of the same dimension. The dimension of invariant subspaces is expressed through the algebra index: dim$M_{(s)}=$ dim$\mathfrak{G}-$ind$\mathfrak{G}-2s$, where $s=0,\ldots,($dim$\mathfrak{G}-$ind$\mathfrak{G})/2$.

Let $F_{\alpha}^{(s)}=F_{\alpha}^{(s)}(f)$ be functions on $\mathfrak{G^{*}}$ that define invariant subspaces $M_{(s)}$ on $\mathfrak{G^{*}}$:
\[
M_{(s)}=\left\{ f\in\mathfrak{G}^{*}\vert F_{\alpha}^{(s)}(f)=0,\quad\neg\left(F_{\alpha}^{(s+1)}(f)=0\right);\quad\alpha=0,\dots,\dim\mathfrak{G}-r_{(s)}\right\} ,
\]
where $r_{(s)}$ is the dimension of the invariant subspaces $M_{(s)}$.

We define an embedding $f:O_{\lambda}\rightarrow\mathfrak{G}^{*}$ that associates each linear functional with its canonical coordinates $(p,q)$ on the orbit. This mapping is uniquely determined by the functions $f_{X}(p,q,\lambda)$, which satisfy the system of equations \cite{darbu}:
\[
\{f_{X},f_{Y}\}=f_{[X,Y]},\quad f_{X}(0,0,\lambda)=\langle\lambda,X\rangle,\quad F_{\alpha}^{(s)}(f_{X})=0,\quad X,Y\in\mathfrak{G}.
\]

We proceed to the complex extension of the Lie algebra $\mathfrak{G}$ and consider canonical embeddings linear in the variables $p$,
\begin{equation}
	f_{X}(p,q,\lambda)=\alpha_{X}^{a}(q)p_{a}+\chi_{X}(q,\lambda),\quad X\in\mathfrak{G}_{\mathbb{C}},\quad a=1,\ldots,\dim Q.\label{kanon}
\end{equation}
The domain of the variables $q$ and $p$ is determined from the condition that the functions are real (\ref{kanon}). For the existence of (\ref{kanon}) it is necessary and sufficient that the linear functional $\lambda$ admits the polarization $\mathfrak{n}\subset\mathfrak{G}_{\mathbb{C}}$ \cite{10_Kirill}. Recall that the polarization $\mathfrak{n}$ is an isotropy subalgebra and the set $Q$ can be represented as a local homogeneous space: $Q\simeq G_{\mathbb{C}}/\exp\mathfrak{n}$. The operators
\begin{equation}
	\ell_{X}(q,\lambda):=if_{X}(-i\partial_{q},q,\lambda)\label{lambda}
\end{equation}
form an irreducible representation of the Lie algebra $\mathfrak{G}$ in the space of smooth functions $L_{2}(Q,\mathfrak{n},\lambda)$, which is called the $\lambda$-representation of $\mathfrak{G}$. Moreover, the Casimir operators in the $\lambda$-representation are functions:
\[
K_{\mu}(-i\ell(q,\lambda))=\kappa_{\mu}(\lambda).
\]

The functions $\alpha_{X}^{a}(q)$ are components of the generators of the representation of the group $G$ on the space $Q$. In this case, the functions $\chi_{X}(q,\lambda)$ are expressed in terms of left-invariant fields on the group $G$,
\[
\chi_{X}(q,\lambda)=\xi_{X}^{\alpha}(s(q))\lambda_{\alpha},
\]
where $s:Q\rightarrow G$ is the canonical section of the bundle $(G,\pi,Q,\exp\mathfrak{n})$ (see, for example, \cite{br14}).

On the manifold $Q$, we introduce the measure $d\mu(q)$ and the scalar product
\begin{equation}
	(\psi_{1},\psi_{2})=\int\limits_{Q}\overline{\psi_{1}(q)}\psi_{2}(q)d\mu(q),\quad\psi_{1},\psi_{2}\in L_{2}(Q,\mathfrak{n},\lambda).\label{scapr}
\end{equation}
The measure $d\mu(q)$ is chosen under the condition that the operators (\ref{lambda}) are self-adjoint with respect to the scalar product (\ref{scapr}). This measure can be determined by the generalized delta function $\delta(q,q')$, which is defined as the solution of the system of equations
\[
\left[ \ell_X(q,\lambda) + \overline{\ell_X(q',\lambda)} \right] \delta(q,q') = 0,\quad
\int_Q \delta(q,q') d\mu(q) = 1.
\]

The $\lambda$-representation of the Lie algebra $\mathfrak{G}$ defined above can be lifted to a unitary representation of the group on $L_{2}(Q,\mathfrak{n},\lambda)$,
\begin{equation}
	\frac{d}{dt}\left.\left(T_{\exp tX}^{\lambda}\right)\right|{}_{t=0}\psi(q)=\ell_{X}(q,\lambda)\psi(q),~\psi\in L_{2}(Q,\mathfrak{n},\lambda).\label{Lgroup}
\end{equation}
The action of the operators $T_{g}^{\lambda}$ is represented in the integral form:
\[
T_{g}^{\lambda}\psi(q)=\int\limits_{Q}\mathcal{\mathscr{D}}_{q\overline{q'}}^{\lambda}(g)\psi(q')d\mu(q'),
\]
where $\mathcal{\mathscr{D}}_{q\overline{q'}}^{\lambda}(g)$ are the generalized kernels of the $\lambda$-representation of the Lie group, which have properties
\begin{eqnarray}
	\mathscr{D}_{q\overline{q'}}^{\lambda}(g_{1}g_{2})& =&\int_{Q}\mathcal{\mathscr{D}}_{q\overline{q''}} ^{\lambda}(g_{1})\mathcal{\mathscr{D}}_{q''\overline{q'}}^{\lambda}(g_{2})d\mu(q''),\nonumber\\
	\mathcal{ \mathscr{D}}_{q\overline{q'}}^{\lambda}(g)&=&\overline{\mathcal{\mathscr{D}}_{q'\overline{q}}^{\lambda}(g^{-1})},\quad\mathcal{\mathscr{D}}_{q\overline{q'}}^{\lambda}(e)=\delta(q,\overline{q'}),\label{condD}
\end{eqnarray}
where $g_{1},g_{2}\in G$; $\delta(q,q')$ is the generalized Dirac delta function with respect to the measure $d\mu(q)$. Generalized functions (\ref{condD}) satisfy the system of equations
\begin{gather}
	\left[\xi_{X}(g)+\ell_{X}(q,\lambda)\right]\mathcal{\mathscr{D}}_{q\overline{q'}}^{\lambda}(g^{-1})=0,\quad\left[\eta_{X}(g)+\overline{\ell_{X}(q',\lambda)}\right]\mathcal{\mathscr{D}}_{q\overline{q'}}^{\lambda}(g^{-1})=0.\label{tD}
\end{gather}
Note that the functions $\mathcal{\mathscr{D}}_{q\overline{q'}}^{\lambda}(g)$ are defined globally on the Lie group $G$ if and only if the Kirillov condition for the orbit $\mathcal{O}_{\lambda}$ to be integral is satisfied:
\begin{equation}
	\frac{1}{2\pi}\oint_{\gamma\in H_{1}(\mathcal{O}_{\lambda})}\omega_{\lambda}=n_{\gamma}\in\mathbb{Z},\label{cel_D}
\end{equation}
where $H_{1}(\mathcal{O}_{\lambda})$ is the one-dimensional homology group of the stationarity group
\[
G^{\lambda}=\{g\in G\mid Ad_{g}^{*}\lambda=\lambda\}.
\]
Generalized functions $\mathcal{\mathscr{D}}_{q\overline{q'}}^{\lambda}(g)$ can be represented as follows:
\begin{equation}
	\mathcal{\mathscr{D}}_{q\overline{q'}}^{\lambda}(g^{-1})=U^{\lambda}(h(q,g^{-1}))\delta(q\circ g^{-1},\overline{q'}),\label{th1}
\end{equation}
where $\delta(q,q')$ is the generalized Dirac delta function with respect to the scalar product (\ref{scapr}), $U^{\lambda}$ is a one-dimensional, unitary irreducible representation of the subgroup $\exp\mathfrak{n}$ in the space $L_{2}(Q,\mathfrak{n},\lambda)$ with the scalar product (\ref{scapr}),
\begin{equation}
	U^{\lambda}(h)=\exp\left(-i\int\lambda_{\beta}\sigma^{\beta}(h)\right),\label{eq21}
\end{equation}
where $\sigma^{\beta}(h):=\sigma^{e_{\beta}}(h)$ is the right-invariant Maurer-Cartan form corresponding to the basis vectors $e_{\beta}$ of the subalgebra $\mathfrak{n}$ .

In the paper \cite{9_Sh_Shir}, it is conjectured that this set of generalized functions has the properties of completeness and orthogonality for a certain choice of the measure $d\mu(\lambda)$:
\begin{gather}
	\int_{G}\overline{\mathcal{\mathscr{D}}_{\tilde{q}\overline{\tilde{q}'}}^{\lambda}(g)}\mathcal{\mathscr{D}}_{q\overline{q'}}^{\lambda}(g)d\mu(g)=\delta(q,\tilde{q})\delta(\tilde{q}',q')\delta(\tilde{\lambda},\lambda),\label{Dort2}\\
	\int_{Q\times Q\times J}\overline{\mathcal{\mathscr{D}}_{q\overline{q'}}^{\lambda}(\tilde{g})}\mathcal{\mathscr{D}}_{q\overline{q'}}^{\lambda}(g)d\mu(q)d\mu(q')d\mu(\lambda)=\delta(\tilde{g},g),\label{Dful2}
\end{gather}
where $\delta(g)$ is the generalized Dirac delta function with respect to the Haar measure $d\mu(g)$ on the Lie group $G$. For compact Lie groups, relations (\ref{Dort2})--(\ref{Dful2}) hold due to the Peter-Weyl theorem \cite{2_Bar_Ron}. The domain $J$ of variation of the parameters $\lambda$ and the measure $d\mu(\lambda)$ are chosen in such a way that relations (\ref{Dort2})--(\ref{Dful2}) are satisfied.

Note that although there is no rigorous proof of the relations (\ref{Dort2})--(\ref{Dful2}), in each specific case it is not difficult to verify their validity by direct calculation.

\section{Non-commutative reduction}

Consider the Schr\"{o}dinger equation for a quantum system with Hamiltonian (\ref{1.3}),
\begin{equation}
	\mathscr{H}\varphi(g)=E\varphi(g),\quad\varphi\in\mathfrak{H}.\label{5.1}
\end{equation}
We find the complete system of solutions to equation (\ref{5.1}) using the method of non-commutative integration.

The basic idea of the non-commutative reduction of equation (\ref{5.1}) is to find a complete set of solutions in the form:
\begin{equation}
	\varphi_{q}^{\lambda}(g^{-1})=\int\limits_{Q}\psi(q',\lambda)\mathcal{\mathscr{D}}_{q\overline{q'}}^{\lambda}(g^{-1})d\mu(q').\label{anzaz}
\end{equation}

Here, the complete set of solutions (\ref{anzaz}) is parameterized by quantum numbers $\lambda$, which correspond to commuting symmetry operators of equation (\ref{5.1}), namely, the Casimir operators of the algebra $\mathfrak{G}$. The set of parameters $q$ takes continuous values.

Let us consider the action of the operators $\eta_{X}(g)$ on the functions (\ref{anzaz}):
\begin{equation}
	\begin{split}\eta_{X}(g)\varphi_{q}^{\lambda}(g^{-1}) & =\int\limits_{Q}\psi(q',\lambda)\eta_{i}(g)\mathcal{\mathscr{D}}_{q\overline{q'}}^{\lambda}(g^{-1})d\mu(q')\\
		&=\int\limits_{Q}\psi(q',\lambda)\left[-\overline{\ell_{X}(q',\lambda)}\mathcal{\mathscr{D}}_{q\overline{q'}}^{\lambda}(g^{-1})\right]d\mu(q')\\ 
		& =\int\limits_{Q}\left[\ell_{X}(q',\lambda)\psi(q',\lambda)\right]\mathcal{\mathscr{D}}_{q\overline{q'}}^{\lambda}(g^{-1})d\mu(q').
	\end{split}
	\label{5.2}
\end{equation}
Here we used the relations (\ref{tD}) and the self-adjointness of the operators $\ell_{X}(q,\lambda)$ with respect to the scalar product (\ref{scapr}). Substituting (\ref{anzaz}) into (\ref{5.1}) and taking into account (\ref{5.2}), we obtain the reduced differential equation:
\begin{equation}
	\left[c^{AB}\ell_{A}(q',\lambda)\ell_{B}(q',\lambda)+c^{A}\ell_{A}(q',\lambda)-E\right]\psi(q',\lambda)=0\,.\label{5.3}
\end{equation}

Formula (\ref{th1}) gives an idea of the general form of solutions of equations on groups (\ref{5.1}). Namely, by virtue of (\ref{anzaz}) we have:
\begin{equation}
	\begin{split}\varphi_{q}^{\lambda}(g^{-1}) & =\int_{Q}\mathcal{\mathscr{D}}_{q\overline{q'}}^{\lambda}(g^{-1})\psi(q',\lambda)d\mu(q')\\ 
		& =\int_{Q}U^{\lambda}(h(q,g^{-1}))\delta(q\circ g^{-1},\overline{q'})\psi(q',\lambda)d\mu(q')\\ 
		& =U^{\lambda}(h(q,g^{-1}))\psi(q\circ g^{-1},\lambda)\,,
	\end{split}
	\label{sol_G}
\end{equation}
where the function $\psi(q',\lambda)$ is a solution to equation (\ref{5.3}).

Using (\ref{th1}), it is easy to obtain an explicit expression for the action of the Lie group $G$ on the functions (\ref{sol_G}):
\begin{align}
	\varphi_{q\circ\tilde{g}^{-1}}^{\lambda}(g^{-1}) & =\int_{Q}U^{\lambda}(h(q\circ\tilde{g}^{-1},g^{-1}))\delta(q\circ(\tilde{g}^{-1}g^{-1}))\psi(q',\lambda)d\mu(q')\nonumber \\
	& =\int_{Q}\overline{U^{\lambda}(h(q,\tilde{g}^{-1}))}U^{\lambda}(h(q,\tilde{g}^{- 1}g^{-1}))\delta(q\circ(\tilde{g}^{-1}g^{-1}))\psi(q,q',\lambda)d\mu(q')\nonumber \\ 
	& =\overline{U^{\lambda}(h(q,\tilde{g}^{-1}))}\varphi_{q}^{\lambda}(\tilde{g}^{-1}g^{-1})=\overline{U^{\lambda}(h(q,\tilde{g}^{-1}))}T^{L}(\tilde{g})\varphi_{q}^{\lambda}(g^{-1})\,.\label{coh_G}
\end{align}

Comparing (\ref{coh_G}) with (\ref{Perelomov_coh}), we get that if
\begin{equation}
	\left|U^{\lambda}(h(q,\tilde{g}^{-1}))\right|^{2}=1,\quad q\in Q,~\quad\tilde{g}\in G,\label{eq30}
\end{equation}
then the states (\ref{coh_G}) are generalized coherent states.

Note that if the polarization $\mathfrak{n}$ is real, then the integrand in (\ref{eq21}) is real and property (\ref{eq30}) is automatically satisfied. Thus, we have shown that if the polarization $\mathfrak{n}$ of the functional $\lambda$ is real, the states (\ref{anzaz}), which are obtained using the method of non-commutative integration, are a special case of generalized Perelomov coherent states. 

\section{Coherent states of the rotation group $SO(3)$}

Consider the three-dimensional rotation group $G=\mathbb{SO}(3)$. We fix some basis $\{e_{a}\}$ of the Lie algebra $\mathfrak{so}(3)$ and introduce canonical coordinates $(\phi,\theta,\psi)\in G$ of the second kind:
\begin{equation}
	g(\phi,\theta,\psi)=e^{\psi e_{3}}e^{\left(\theta-\frac{\pi}{2}\right)e_{2}}e^{\phi e_{3}},\quad\phi\in(0;2\pi],\quad\theta\in(0;\pi],\quad\psi\in(0;2\pi].\label{cc2}
\end{equation}
Left-invariant and right-invariant vector fields on the group $\mathbb{SO}(3)$ in canonical coordinates (\ref{cc2}) have the form: 
\begin{eqnarray*}
	\xi_{1}&=&\partial_{\phi},\quad\xi_{2}=-\cot\theta\sin\phi\partial_{\phi}+\cos\phi\partial_{\theta}+\frac{\sin\phi}{\sin\theta}\partial_{\psi}\,,\\
	\xi_{3}&=&-\cot\theta\cos\phi\partial_{\phi}-\sin\phi\partial_{\theta}+\frac{\cos\phi}{\sin\theta}\partial_{\psi},\\
	\eta_{1}&=&-\frac{\cos\psi}{\sin\theta}\partial_{\phi}+\sin\psi\partial_{\theta}+\cos\psi\cot\theta\partial_{\psi}\,,\\
	\eta_{2}&=&-\frac{\sin\psi}{\sin\theta}\partial_{\phi}-\cos\psi\partial_{\theta}+\sin\psi\cot\theta\partial_{\psi},\quad \eta_{3}=-\partial_{\psi}\,.
\end{eqnarray*}
With respect to the coordinate system (\ref{cc2}), the Haar measure $d\mu(g)$ can be given in the following form:
\begin{equation}
	\int_{\mathbb{SO}(3)}(\cdot)d\mu(g)=\frac{1}{8\pi^{2}}\int_{0}^{2\pi}d\psi\int_{0}^{\pi}\sin\theta d\theta\int_{0}^{2\pi}d\phi(\cdot).\label{}
\end{equation}

Each non-degenerate integer C-orbit of the group $\mathbb{SO}(3)$ passes through the covector $\lambda(j)=(j,0,0)$, where $j=1,2,3,\dots$ and is a two-dimensional sphere, $\mathcal{O_{\lambda}}=\left\{ f\in\mathbb{R}^{3}\left|K(f)=j^{2},f\neq0\right.\right\}$. Here $K(f)=f_{1}^{2}+f_{2}^{2}+f_{3}^{2}$ -- the Casimir function of the algebra $\mathfrak{so}(3)$. The complex polarization $\mathfrak{n}=\{e_{1},e_{2}+ie_{3}\}$ of the covector $\lambda(j)$ corresponds to the operators of the $\lambda$-representation
\begin{equation}
	\ell_{1}(q,j)=-i\left[\sin(q)\partial_{q}-j\cos(q)\right]\quad\ell_{2}(q,j)=-i\left[\cos(q)\partial_{q}+j\sin(q)\right],\quad\ell_{3}(q,j)=\partial_{q}.\label{lprSO3-1}
\end{equation}
The operators $-i\ell_{X}(q,j)$ are Hermitian with respect to the scalar product
\[
(\psi_{1},\psi_{2})=\int_{Q}\overline{\psi_{1}(q)}\psi_{2}(q)d\mu_{j}(q),\quad d\mu_{j}(q)=\frac{(2j+1)!}{2^{j}(j!)^{2}}\frac{dq\wedge d\overline{q}}{\left[1+\cos(q-\overline{q})\right]^{j+1}}.
\]
The functions $\mathcal{\mathscr{D}}_{q\overline{q'}}^{\lambda}(g^{-1})=\langle\phi,\theta,\psi\mid j,q,q'\rangle$ have the form
\begin{gather}
	\mathcal{\mathscr{D}}_{q\overline{q}'}^{j}(\phi,\theta,\psi)=\frac{2^{j}(j!)^{2}}{(2j)!}\left[\sin\theta\cos\phi+\cos\left(\psi+\overline{q}'\right)(\cos q\cos\phi-i\sin\phi)-i\sin\theta\cos q\sin\phi\right.\label{Dso3}\\
	\left.-i\cos\theta\sin q+\sin\left(\psi+\overline{q}'\right)(-i\cos\theta\cos\phi-\cos\theta\cos q\sin\phi+\sin\theta\sin q)\right]{}^{j}.\nonumber 
\end{gather}
With respect to the measure $d\mu(\lambda)$ and the delta function $\delta_{j}(q,\overline{q'})$ of the form
\begin{equation}
	\int_{J}(\cdot)d\mu(\lambda)=\sum_{j=0}^{\infty}(2j+1)(\cdot),\quad\delta_{j}(q,\overline{q'})=\frac{2^{j}(j!)^{2}}{(2j)!}\left[1+\cos(q-\overline{q'})\right],\label{-1}
\end{equation}
the conditions of completeness and orthogonality (\ref{Dort2})--(\ref{Dful2}) on generalized functions (\ref{Dso3}) are satisfied. The right action $q'=q\circ g^{-1}$ of the group $\mathbb{SO}(3)$ on $Q$ is given by the expression
\[
\tan\frac{q'}{2}=\frac{\alpha(\phi,\theta-\frac{\pi}{2},\psi)\tan\frac{q}{2}+\beta(\phi,\theta-\frac{\pi}{2},\psi)}{-\beta^{*}(\phi,\theta-\frac{\pi}{2},\psi)\tan\frac{q}{2}+\alpha^{*}(\phi,\theta-\frac{\pi}{2},\psi)}\text{,}
\]
where
\[
\begin{pmatrix}\alpha(\phi,\theta,\psi) & \beta(\phi,\theta,\psi)\\
	-\beta^{*}(\phi,\theta,\psi) & \alpha^{*}(\phi,\theta,\psi)
\end{pmatrix}=\begin{pmatrix}\cos\frac{\psi}{2} & -\sin\frac{\psi}{2}\\
	\sin\frac{\psi}{2} & \cos\frac{\psi}{2}
\end{pmatrix}\begin{pmatrix}\cos\frac{\theta}{2} & i\sin\frac{\theta}{2}\\
	i\sin\frac{\theta}{2} & \cos\frac{\theta}{2}
\end{pmatrix}\begin{pmatrix}e^{i\phi/2} & 0\\
	0 & e^{-i\phi/2}
\end{pmatrix}.
\]

Let us introduce a common eigenfunction for the maximal commutative set of the enveloping algebra of left-invariant and right-invariant vector fields of the group $\mathbb{SO}(3)$:
\begin{align}
	& K(-i\xi)\mathcal{D}_{mn}^{j}(g)=j(j+1)\mathcal{D}_{mn}^{j}(g),\nonumber \\
	& -i\xi_{1}\mathcal{D}_{mn}^{j}(g)=m\mathcal{D}_{mn}^{j}(g),\nonumber \\
	& i\eta_{3}\mathcal{D}_{mn}^{j}(g)=n\mathcal{D}_{mn}^{j}(g).\label{DmnjEq}
\end{align}
The solution to this system of equations (\ref{DmnjEq}) within the framework of the method of separation of variables is:
\begin{align}
	& \mathcal{D}_{mn}^{j}(g)=\mathcal{D}_{mn}^{j}(\phi,\theta,\psi)=\langle\phi,\theta,\psi\mid j,m,n\rangle=e^{im\phi+in\psi}d_{mn}^{j}(\theta),\label{dmn1}
\end{align}
\begin{align*}
	& d_{mn}^{j}(\theta)=(-1)^{m-n}\sqrt{\frac{(j+m)!(j-m)!}{(j+n)!(j-n)!}}\sin^{m-n}\frac{\theta}{2}\cos^{m+n}\frac{\theta}{2}P_{j-m}^{(m-n,m+n)}(\cos\theta),\\
	& m=-j,\dots,j;\quad n=-j,\dots,j.
\end{align*}
Here $P_{n}^{(\alpha,\beta)}(z)$ are Jacobi polynomials, $j=1,2,\dots$,
\[
P_{n}^{(\alpha,\beta)}(z)=\frac{(-1)^{n}}{2^{n}n!}(1-z)^{-\alpha}(1+z)^{-\beta}\frac{d^{n}}{dz^{n}}\left[(1-z)^{n+\alpha}(1+z)^{n+\beta}\right]\text{.}
\]
The set $\mid j,m,n\rangle$ forms an irreducible unitary $(2j+1)$-dimensional representation of the Lie group $\mathbb{SO}(3)$ with weight $j$. The functions $\mathcal{D}_{mn}^{j}(g)$ are a matrix of finite rotations. These functions satisfy the completeness and orthogonality conditions:
\begin{gather}
	\int_{G}\overline{\mathcal{D}_{mn}^{j}(g)}\mathcal{D}_{\tilde{m}\tilde{n}}^{\tilde{j}}(g)d\mu(g)=\frac{\delta_{j\tilde{j}}}{2j+1}\delta_{m\tilde{m}}\delta_{n\tilde{n}},\label{Dvig1}\\
	\sum_{n=-j}^{j}\overline{\mathcal{D}_{mn}^{j}(g)}\mathcal{D}_{\tilde{m}n}^{j}=\delta_{m\tilde{m}}.\label{Dvig2}
\end{gather}

Let's apply the method of non-commutative integration to the system of equations (\ref{DmnjEq}). We will seek a solution in the form
\[
\mathcal{D}_{mn}^{j}(\phi,\theta,\psi)=C_{mn}^{j}\int_{Q\times Q}\overline{F_{m}^{j}(q)}\Phi_{n}^{j}(q')\mathcal{\mathscr{D}}_{q\overline{q}'}^{j}(\phi,\theta,\psi)d\mu_{j}(q)d\mu_{j}(q').
\]
Then we obtain reduced equations for the functions $F_{m}^{j}(\overline{q})$ and $\Phi_{n}^{j}(q')$:
\[
i\ell_{1}(q,j)F_{m}^{j}(q)=mF_{m}^{j}(q),\quad i\ell_{3}(q',j)\Phi_{n}^{j}(q')=n\Phi_{n}^{j}(q').
\]
We will choose particular solutions in the form:
\[
F_{m}^{j}(q)=\tan^{m}\left(\frac{q}{2}\right)\sin^{j}q,\quad\Phi_{n}^{j}(q')=e^{-inq'}.
\]
The normalization factor $C_{mn}^{j}$ is determined by the expression
\[
C_{mn}^{j}=\exp\left[i\frac{\pi}{2}(j+m)\right](j!)^{2}\left[(j+m)!(j-m)!(j+n)!(j-n)!\right]^{-1/2}.
\]
Thus, we have obtained a connection between the functions $\mathcal{\mathscr{D}}_{q\overline{q}'}^{j}(\phi,\theta,\psi)$ and the Wigner function $\mathcal{D}_{mn}^{j}(\phi,\theta,\psi)$, which is natural to write in invariant notation:
\begin{equation}
	\mid j,m,n\rangle=C_{mn}^{j}\int{}_{Q\times Q}\overline{\tan^{m}\left(\frac{q}{2}\right)\sin^{j}q}e^{-inq'}d\mu_{j}(q)d\mu_{j}(q')\mid j,q,q'\rangle.\label{rel1}
\end{equation}

Let us consider a system of equations on the group $\mathbb{SO}(3)$ of the following form:
\begin{align}
	& K(-i\xi)\Phi(g)=j(j+1)\Phi(g)\,,\nonumber \\
	& -i\eta_{3}\Phi(g)=0\,.\label{eqsys1}
\end{align}
Within the framework of the non-commutative integration method, the basis of solutions of the system of equations (\ref{eqsys1}) is determined by functions of the form:
\begin{align*}
	\Phi(g)=\mathscr{\mathcal{\mathscr{D}}}_{q}^{j}(\phi,\theta) & =\int_{Q}\mathcal{\mathscr{D}}_{q\overline{q}'}^{j}(\phi,\theta,\psi)d\mu_{j}(q')=\\
	& =\left[-i\cos\theta\sin q+\left(\cos\phi-i\cos q\sin\phi\right)\sin\theta\right]^{j}.
\end{align*}

The group $\mathbb{SO}(3)$ acts transitively by right shifts on the two-dimensional sphere $\mathbb{S}^{2}\approx\mathbb{SO}(3)/\mathbb{SO}(2)$, which is a symmetric homogeneous space with coordinates $(\phi,\theta)$ and isotropy subgroup $\mathbb{SO}(2)=\exp(\psi e_{3})$,
\begin{align*}
	& \mathbb{SO}(3)=\left(\mathbb{SO}(3),\pi,\mathbb{SO}(2)\right),\\
	& \pi(g)=x,\quad g=(\phi,\theta,\psi)\in\mathbb{SO}(3),\quad x=(\phi,\theta)\in\mathbb{S}^{2},\quad h=(\psi)\in\mathbb{SO}(2).
\end{align*}
We choose the section $s(x)=g(\phi,\theta,0)$, $x\circ g=\pi\left[s(x)\circ g\right]$. The generators of the group action on the sphere have the form
\begin{align*}
	X_{1} & =\partial_{\phi},\\
	X_{2} & =-\cot\theta\sin\phi\partial_{\phi}+\cos\phi\partial_{\theta},\\
	X_{3} & =-\cot\theta\cos\phi\partial_{\phi}-\sin\phi\partial_{\theta}.
\end{align*}

Spherical functions
\[
Y_{m}^{j}(\phi,\theta)=\sqrt{\frac{2j+1}{4\pi}}\mathcal{D}_{m0}^{j}(\phi,\theta,\psi)
\]
represent common eigenfunctions of the Casimir operator $K(-iX)$ and the operator $-iX_{1}$:
\begin{align*}
	& K(-iX)Y_{m}^{j}(\phi,\theta)=j(j+1)Y_{m}^{j}(\phi,\theta),\\
	& -iX_{1}Y_{m}^{j}(\phi,\theta)=mY_{m}^{j}(\phi,\theta).
\end{align*}
Moreover, the angular momentum operators are related to the operators of the representation of the group $\mathbb{SO}(3)$ on $\mathbb{S}^{2}$:
\begin{align*}
	& \hat{L}_{x}=-i\hbar X_{3},\quad\hat{L}_{y}=-i\hbar X_{2},\quad\hat{L}_{z}=-i\hbar X_{1}.\\
	& \hat{L}^{2}=K(-i\hbar X).
\end{align*}
From (\ref{rel1}) it follows
\begin{align}
	Y_{m}^{j}(\phi,\theta) & =\sqrt{\frac{2j+1}{4\pi}}C_{mn}^{j}\int{}_{Q}\overline{\tan^{m}\left(\frac{q}{2}\right)\sin^{j}q}\left[\int{}_{Q}\mathcal{\mathscr{D}}_{q\overline{q}'}^{j}(\phi,\theta,\psi)d\mu_{j}(q')\right]d\mu_{j}(q)\nonumber \\
	& =\sqrt{\frac{2j+1}{4\pi}}\frac{2^{j}j!}{\sqrt{(j-m)!(j+m)!}}\int_{Q}\frac{\left(i\tan\frac{\overline{q}}{2}\right)^{j+m}}{1-\left(i\tan\frac{\overline{q}}{2}\right)^{2}}\mathcal{\mathscr{D}}_{q}^{j}(\phi,\theta)d\mu_{j}(q).\label{rel2}
\end{align}
The orthogonality of the spherical functions
\begin{align*}
	& \int_{\mathbb{S}^{2}}\overline{Y_{m}^{j}(\phi,\theta)}Y_{m'}^{j'}(\phi,\theta)d\mu(x)=\delta_{jj'}\delta_{mm'},\\
	& \int_{\mathbb{S}^{2}}(\cdot)d\mu(x)=\int_{0}^{2\pi}d\phi\int_{0}^{\pi}\sin\theta d\theta(\cdot),
\end{align*}
implies the orthogonality of the functions $\mathscr{D}_{q}^{j}(\phi,\theta)$:
\[
\int_{\mathbb{S}^{2}}\overline{\mathcal{\mathscr{D}}_{q}^{j}(\phi,\theta)}\mathcal{\mathscr{D}}_{q'}^{j'}(\phi,\theta)\frac{d\mu(x)}{4\pi}=\delta_{j}\left(\overline{q},q'\right)\delta_{j,j'}.
\]

Let $\mid j,m\rangle$ denote the basis of the representation of the group $\mathbb{SO}(3)$ on the sphere $\mathbb{S}^{2}$, which in the coordinate representation is described by the spherical functions $Y_{m}^{j}(\phi,\theta)=\langle\phi,\theta\mid j,m\rangle$. Similarly, let the set $\mid q,j\rangle$ correspond to the functions $\mathscr{\mathcal{D}}_{q}^{j}(\phi,\theta)=\langle\phi,\theta\mid q,j\rangle$ in the coordinate representation. Then expression (\ref{rel2}) and its inverse can be written in invariant form:
\begin{align*}
	& \mid j,m\rangle=\int_{Q}d\mu_{j}(q)\langle q,j\mid j,m\rangle\mid q,j\rangle,\\
	& \mid q,j\rangle=\sum_{m=-j}^{j}\langle j,m\mid q,j\rangle\mid j,m\rangle,\\
	& \langle q,j\mid j,m\rangle=\sqrt{\frac{2j+1}{4\pi}}\frac{2^{j}j!}{\sqrt{(j-m)!(j+m)!}}\left(i\tan\frac{\overline{q}}{2}\right)^{j+m}\left[1-\left(i\tan\frac{\overline{q}}{2}\right)^{2}\right]^{-j}.
\end{align*}

The expansion of spin coherent states (CS) of the group $\mathbb{SO}(3)$ in the orthonormal basis $\mid j,m\rangle$ has the form \cite{1_Charp, 8_perelomov}:
\begin{align*}
	& \mid\zeta,j\rangle=\sum_{m=-j}^{j}u_{m}(j)\mid j,m\rangle,\\
	& u_{m}(j)=\left(1+\left|\zeta\right|^{2}\right)^{-j}\sqrt{\frac{(2j)!}{(j+m)!(j-m)!}}\zeta^{j+m},\\
	& \zeta=-\tan\frac{\overline{\theta}}{2}\exp(-i\overline{\phi}),\quad(\overline{\phi},\overline{\theta})\in S^{2}.
\end{align*}
In coordinate representation we have:
\begin{align*}
	\psi_{\zeta}^{j}(\phi,\theta) & =\langle x\mid\zeta,j\rangle=\\
	& =\sqrt{\frac{2j+1}{4\pi}}\frac{\sqrt{(2j)!}}{2^{j}j!}\left(1+\left|\zeta\right|^{2}\right)e^{-ij\phi}\left(2e^{i\phi}\cos\theta\cdot\zeta+\sin\theta-e^{2i\phi}\sin\theta\cdot\zeta^{2}\right)^{j}.
\end{align*}
The action of the rotation group in the space of quantum numbers $\zeta$ is defined using a linear-fractional transformation of the form
\[
(\text{i\ensuremath{\zeta)\circ g^{-1}=\frac{\alpha(\phi,\theta-\frac{\pi}{2},\psi)(i\zeta)+\beta(\phi,\theta-\frac{\pi}{2},\psi)}{-\beta^{*}(\phi,\theta-\frac{\pi}{2},\psi)(i\zeta)+\alpha^{*}(\phi,\theta-\frac{\pi}{2},\psi)}},}
\]
Then the main property of coherent states of a rotation group takes place:
\begin{equation}
	\left[R_{g}^{*}\psi_{\zeta}^{j}\right](\phi,\theta)=\psi_{\zeta}^{j}\left(x(\phi,\theta)\circ g\right)=e^{ij\Phi^{j}(\zeta,g)}\psi_{\zeta\circ g^{-1}}^{j}(\phi,\theta),\label{r0}
\end{equation}
where $\Phi^{j}(\zeta,g)$ is the real function.

The relationship between non-commutative states $|q,j\rangle$ and coherent ones takes a simple form:
\begin{equation}
	\mid q,j\rangle=\sqrt{\frac{4\pi}{2j+1}}\frac{2^{j}j!}{\sqrt{(2j)!}}\left\{ \frac{1+\left|-i\tan\frac{q}{2}\right|^{2}}{1-\left(-i\tan\frac{q}{2}\right)^{2}}\right\} \mid-i\tan\frac{q}{2},j\rangle.\label{rel3}
\end{equation}
We note that,
\begin{align}
	& \left[R_{g}^{*}\mathcal{\mathscr{D}}_{q}^{j}\right](\phi,\theta)=\mathcal{\mathscr{D}}_{q}^{j}((\phi,\theta)\circ g)=\mathcal{\mathscr{D}}_{q}^{j}(\pi(g))\mathcal{\mathscr{D}}_{q\circ g^{-1}}^{j}(\phi,\theta).\label{r1}
\end{align}
Comparing (\ref{r0}) and (\ref{r1}) taking into account (\ref{rel3}), we obtain
\[
\mathscr{\mathcal{D}}_{q}^{j}(\pi(g))=e^{ij\Phi^{j}(\zeta,g)}\left(\frac{f(\zeta)}{f(\zeta g^{-1})}\right)^{j},\quad f(\zeta)=\frac{1+\left|\zeta\right|^{2}}{1-\zeta^{2}},\quad\zeta=-i\tan\frac{q}{2}.
\]
It follows from (\ref{r1}) that the non-commutative states $| q,j\rangle$ are not spin CSs, since the modulus of function
\[
\frac{f(\zeta)}{f(\zeta g^{-1})}
\]
varies depending on the group element $g$. However, acting by the group $\mathbb{SO}(3)$ on $Q$ we obtain all non-commutative states from some fixed ket-vector:
\begin{equation}
	\mid q,j\rangle=\mid q_{0}\circ g,j\rangle=\mathcal{\mathscr{D}}_{q_{0}}^{j}\left(\pi[g^{-1}]\right)R_{g^{-1}}\mid q_{0},j\rangle,\quad q_{0}=0.\label{rel4}
\end{equation}
Relation (\ref{rel4}) generalizes the well-known property of coherent states (\ref{r0}) to the case of states that are given by the method of non-commutative integration.

\section{Concluding remarks}

In this paper, we study the Schr\"{o}dinger equation on Lie groups in which the Hamiltonian admits a non-Abelian Lie algebra of left-invariant vector fields on the Lie group.

In this case, the method of non-commutative integration of linear differential equations is effectively used to construct a complete set of solutions. As a result of non-commutative reduction, we arrive at a reduced equation (\ref{5.3}), which depends on a smaller number of independent variables $q'$. The complete set of solutions (\ref{sol_G}) is parameterized by the set of parameters $(q,\lambda)$, where the parameters $\lambda$ are determined by the eigenvalues of the Casimir operators $K_\mu(-i\xi)$,
\begin{equation}
	K_\mu(-i\xi)\varphi_{q}^{\lambda}(g^{-1}) = \kappa_\mu(\lambda) K_\mu(-i\xi).
\end{equation}
Moreover, the parameters $q$, which number the solutions, take continuous values and do not correspond, in general, to commuting integrals of motion. In other words, unlike the real parameters $\lambda$, they are not quantum numbers.

The paper shows that if the Lie algebra $\mathfrak{G}$ admits real polarization, then the set of solutions (\ref{sol_G}) is a special case of  Perelomov coherent states, in which the parameters $q$ label different quantum states.

If the polarization is complex, then the solutions (\ref{sol_G}) are, in a sense, a generalization of Perelomov coherent states. Moreover, the action of the group on the manifold $Q$ changes not only the phase but also the modulus of the wave function (\ref{sol_G}). To illustrate this situation, we obtain a relation (\ref{rel3}) between states obtained using the noncommutative integration method on the rotation group $\mathbb{SO}(3)$ and spin coherent states. The resulting solutions satisfy the group property (\ref{rel4}), which generalizes the property (\ref{r0}) of spin coherent states. We also find a new integral representation (\ref{rel2}) for spherical harmonics.

\section*{Acknowledgements}

D. M. G. thanks FAPESP (Grant No. 21/10128-0) and CNPq for permanent support.

\end{document}